\documentclass[aps,twocolumn,a4paper,floatfix,superscriptaddress,pra]{revtex4-2}
\usepackage[toc,page]{appendix}
\usepackage{braket}
\usepackage{epsfig,graphicx,times}
\usepackage{amstext}
\usepackage{amsmath}
\usepackage{amssymb}
\usepackage{mathrsfs}
\usepackage{dcolumn}
\usepackage{bm}
\usepackage[tight]{subfigure}
\usepackage[colorlinks,linkcolor=blue,anchorcolor=blue,citecolor=blue,urlcolor=blue]{hyperref}
\usepackage{color}
\usepackage{siunitx}
\usepackage{appendix}
\usepackage{placeins}
\usepackage{textcomp}
\usepackage{bbold}
\usepackage{float}
\usepackage{verbatim}
\usepackage{minitoc}
\usepackage[dvipsnames]{xcolor}

\usepackage{soul}
\usepackage[framemethod=tikz]{mdframed}
\usepackage{colortbl}  
\usepackage{xcolor}
\usepackage{array}
\begin{document}

\title{Temperature-enhanced quantum sensing for the cutoff frequency of Ohmic environments}
\author{Ji-Bing Yuan}\email{jbyuan@hynu.edu.cn}
\affiliation{Key Laboratory of Opto-electronic Control and Detection Technology of University of Hunan Province, and College of Physics and Electronic Engineering, Hengyang Normal University, Hengyang 421002, China}

\author{Ya-Ju Song}
\affiliation{Key Laboratory of Opto-electronic Control and Detection Technology of University of Hunan Province, and College of Physics and Electronic Engineering, Hengyang Normal University, Hengyang 421002, China}

\author{Shi-Qing Tang}\email{sqtang@hynu.edu.cn}
\affiliation{Key Laboratory of Opto-electronic Control and Detection Technology of University of Hunan Province, and College of Physics and Electronic Engineering, Hengyang Normal University, Hengyang 421002, China}

\author{Xin-Wen Wang}
\affiliation{Key Laboratory of Opto-electronic Control and Detection Technology of University of Hunan Province, and College of Physics and Electronic Engineering, Hengyang Normal University, Hengyang 421002, China}

\author{Le-Man Kuang}\email{lmkuang@hunnu.edu.cn}
\affiliation{Key Laboratory of Low-Dimensional Quantum Structures and Quantum Control of Ministry of Education, and Department of Physics, Hunan Normal University, Changsha 410081, China}
\affiliation{Synergetic Innovation Academy for Quantum Science and Technology, Zhengzhou University of Light Industry, Zhengzhou 450002, China}
\date{\today}
\begin{abstract}
We investigate the quantum sensing performance of a dephasing qubit as a probe in Ohmic environments, characterized by the coupling strength $\eta$, the Ohmicity parameter $s$, and the cutoff frequency $\omega_c$ to be estimated. The performance is quantified by the dimensionless quantum signal-to-noise ratio $\mathcal{Q}$. We show that the evolution of  $\mathcal{Q}$ with the scaled time $\omega_c t$ is independent of $\omega_c$, and peaks at an optimal time $t_{\text{opt}}$, yielding optimal sensitivity $\mathcal{Q}_{\text{opt}}$. We analyze how $\mathcal{Q}_{\text{opt}}$ depends on $\eta$, $s$ and the temperature $T$.  Our results demonstrate that, for any Ohmic environment, provided that $\omega_c t_{\text{opt}} \ll 1$, $\mathcal{Q}_{\text{opt}}$ always reaches the upper bound: $\mathcal{Q}_{\text{max}} = 0.648$ at zero temperature, and consistently attains $\mathcal{Q}_{\text{max}}/4$ at high temperatures. Remarkably, we find that increasing the scaled temperature $T/\omega_c$ can enhance $\mathcal{Q}_{\text{opt}}$ by nearly two orders of magnitude compared to its zero-temperature counterpart for certain Ohmic environments. Our work reveals that temperature can serve as a resource to enhance sensing precision, as it accelerates the encoding of the cutoff frequency information into the probe state, thereby enabling optimal measurement within a short time window.
\end{abstract}
\maketitle
\section{\label{Sec:1}Introduction}
Quantum sensing~\cite{Cappellaro2017} seeks to harness quantum resources such as coherence~\cite{Adam2022,Ullah2023,Aiachei2024}, entanglement~\cite{Nagata2007,Treutlein2018,Haine2020,Planella2022}, and squeezing~\cite{Muessel2014,Bai2019} to enhance the precision of various physical measurements. Accurate sensing of the environmental parameters surrounding a quantum system is crucial for both theoretical research on quantum open systems and practical applications of quantum technology, as all quantum systems inevitably interact with their environment, leading to decoherence~\cite{Breuer2007}. However, this task remains challenging due to the high dimensionality and complexity of environmental degrees of freedom. A powerful approach to address this issue involves the use of quantum probes~\cite{Montenegro2025,Benedetti2014,Grasselli2018,Mehboudi2019,Song2019,Wu2021,zhou2021,Khan2022,Zhang2022,Aiache2024,Midha2025,Wu2025}. These are small, controllable, and measurable quantum systems that are initially prepared in an appropriate state and then coupled to the environment.  During this interaction, information about the target environmental parameters becomes encoded into the probe's quantum state, which can subsequently be extracted through suitable measurement protocols. The estimation precision has been extensively analyzed within the framework of quantum parameter estimation theory~\cite{Helstrom1976,Holevo1982}. According to this formalism, the ultimate precision limit of any estimation procedure is set by the quantum Cram\'{e}r-Rao bound (QCR), quantitatively characterized via the quantum Fisher information (QFI) or the corresponding dimensionless quantum signal-to-noise ratio (QSNR)~\cite{Wei2013,Jing2019}. A larger QFI (QSNR) corresponds to greater potential accuracy.

Among various candidates, qubit-based quantum probes have attracted significant attention owing to their simplicity and broad applicability in environmental sensing tasks~\cite{Tan2013,Jevtic2015,Porras2017,Tamascelli2020,Chu2020,Wolski2020,Shi2020,Blondeau2021,Ai2021,Zheng2022,Xu2023,Tan2024,Abbas2024,Abbas2025,Dong2025}. To minimize the disturbance of qubits to their environment, researchers often design interaction models in which the qubit do not exchange energy with the surroundings~\cite{Hu2025,Balatsky2025,Cai2025,Yuan2025}. This type of model is referred to as a dephasing model. As a matter of fact, dephasing qubits are widely used to measure various properties of the environment, such as temperature~\cite{Razavian2019,Mitchison2020,Francesca2020,Candeloro2021,Yuan2023,Smirne2023,Brattegard2024}, system-bath coupling strength~\cite{Tan2020,Yuan2024,Xu2025}, the non-Markovian properties~\cite{Haikka2011,Haikka2013,Yuan2017}, and the cutoff frequency of Ohmic environments~\cite{Benedetti2018,Sehdaran2019,Bahrampour2019,Ather2021,Tan2022,Zheng2025}. The Ohmic environment is a typical setting in the study of open quantum systems. Its spectral density function is characterized by three parameters: the cutoff frequency $\omega_{c}$, the Ohmic parameter $s$, and, the dimensionless coupling strength $\eta$. Accurate sensing of the cutoff frequency in an Ohmic environment is crucial because it determines the environmental correlation time, governs decoherence dynamics, and enables precise modeling of open quantum systems. Mastery of this parameter facilitates better noise suppression and discrimination between Markovian and non-Markovian behaviors~\cite{Breuer2007,Zwerger1987}.

Benedetti et al. proposed a scheme for estimating the cutoff frequency of Ohmic environments using single-dephasing qubit and two-dephasing qubit probes~\cite{Benedetti2018}. They pointed out that \textit{"for most values of the Ohmicity parameter, a simple probe such as a single qubit is already optimal for the precise estimation of the cutoff frequency"}. Subsequently, they also investigated the effect of temperature on the sensing performance of the single-dephasing qubit~\cite{Sehdaran2019}. It should be noted that during their study, they fixed the coupling strength $\eta$ to $1$. We found that the dynamics of an impurity atom in a one-dimensional atomic Bose-Einstein condensate can be described by the dephasing model, where the condensate acts as an $s=1$ Ohmic environment. The weak coupling strength ($\eta\ll1$) can be widely tuned by changing the scattering length between the impurity atom and the condensate~\cite{Yuan2023}. Meanwhile, many studies have shown that the coupling strength between the system and the environment significantly affects sensing performance~\cite{Mitchison2020,Yuan2023,Correa2017,Aiache2025,Tan2025}. These works motivate us to investigate utilizing a single dephasing qubit to sense the cutoff frequency of  Ohmic environments with variable coupling strength. In our study, we demonstrate that increasing the coupling strength $\eta$ enhances the optimal QSNR $\mathcal{Q}_{\text{opt}}$ at the optimal measurement time $t_{\text{opt}}$. Notably, when $\omega_c t_{\text{opt}} \ll 1$, the optimal QSNR consistently reaches the upper bound of this sensing scheme at zero-temperature, $\mathcal{Q}_{\text{max}} = 0.648$, regardless of variations in the characteristic parameters of the Ohmic environments. Our results indicate that for Ohmic environments with small coupling strength $\eta$ and low Ohmicity parameter $s$, since the optimal measurement time does not fall within the short-time regime, the achievable $\mathcal{Q}_{\text{opt}}$ is significantly lower than this upper limit. This observation raises an intriguing question: if the effect of temperature is considered, which accelerates the decoherence process through thermal noise, can the optimal measurement time be compressed into a short time interval? If this occurs, what changes would take place in the optimal QSNR? We find that increasing the temperature indeed enables the optimal measurement time to enter the short-time regime. Our observations reveal that, at sufficiently high temperatures, the optimal QSNR in all Ohmic environments approaches $\mathcal{Q}_{\text{max}}/4$. This indicates that, for Ohmic environments with weak coupling $\eta$ and low Ohmicity parameter $s$, temperature becomes a resource for enhancing quantum sensing precision by speeding up information acquisition. These findings offer new perspectives for achieving higher-precision sensing.

The paper is structured as follows: In Sec.~\ref{Sec:2}, we present the quantum sensing protocol. In Sec.~\ref{Sec:3}, We study the sensing performance of the dephasing qubit sensor in estimating the cutoff frequency at zero temperature. The impact of the environment's temperature on sensing performance is discussed in Sec.~\ref{Sec:4}.  Finally, a conclusion is drawn in Sec.~\ref{Sec:5}.

\section{\label{Sec:2} quantum sensing protocol}
We consider a single qubit interacting with an environment consisting of numerous harmonic oscillator modes. The interaction follows the dephasing model. Consequently, the total system's  Hamiltonian is given by
\begin{equation}
\label{hami}
 \hat{H}=\frac{\omega_{0}}{2}\hat{\sigma}_{z}+\sum_{{\bf k}}\omega_{{\bf
k}}\hat{b}^{\dag}_{{\bf k}}\hat{b}_{{\bf k}}+\sum_{\mathbf{k}}\hat{\sigma}_{z}(g_{\mathbf{k}}\hat{b}^{\dag}_{\mathbf{k}}+g^{*}_{\mathbf{k}}\hat{b}_{\mathbf{k}}),
\end{equation}
where $\omega_{0}$ and $\hat{\sigma}_{z}$ are the level splitting and the Pauli-Z operator of the qubit, $\hat{b}_{\mathbf{k}}$ ($\hat{b}_{\mathbf{k}}^{\dag}$) is bosonic annihilation operator (creation operator) for mode $\bf k$,  satisfying the commutation relation $[\hat{b}_{\mathbf{k}},\hat{b}_{\mathbf{k}}^{\dag}]=1$, $\omega_{{\bf k}}$ is the frequency of the $\mathbf{k}$-th mode, $g_{\mathbf{k}}$ denotes the coupling strength between the qubit and the $\mathbf{k}$-th environment mode. Hereafter we set the reduced Planck constant $\hbar=1$.

The initial state of the total system is prepared as a tensor product
\begin{equation}
\hat{\rho}_{tot}(0)=|\psi\rangle\langle\psi|\otimes\hat{\rho}_{E} \nonumber,
\end{equation}
where $|\psi\rangle=1/\sqrt{2}(|0\rangle+|1\rangle)$ is the superposition
state of the qubit with $|0\rangle$ and  $|1\rangle$ being the eigenstates of $\hat{\sigma}_{z}$, $\hat{\rho}_{E}=\prod_{\mathbf k}\left(1-\mathrm{e}^{-\omega_{k}/T}\right)\mathrm{e}^{-\omega_{k}b_{{\mathbf k}}^{\dag }b_{{\bf k}}/T}$ is a thermal state of the environment with $T$ being the temperature.  Hereafter we set the Boltzmann constant $k_{B}=1$. In the dephasing model~(\ref{hami}), the quantum state of the qubit evolving over time $t$ can be precisely given~\cite{Breuer2007,Zwerger1987}. Its density matrix in the computational basis $\{|1\rangle, |0\rangle\}$ takes the following form
\begin{eqnarray}\label{sta1}
\hat{\rho}(t)= \frac{1}{2}\left(
       \begin{array}{cc}
        1 & e^{i\omega_{0}t}e^{-\gamma(t)} \\
         e^{-i\omega_{0}t}e^{-\gamma(t)} &  1\\
       \end{array} \right).
\end{eqnarray}
Here the decoherence factor $\gamma$ is given as
\begin{eqnarray}\label{garm1}
\gamma=\int_{0}^{\infty} d\omega \frac{J(\omega)}{\omega^{2}}\left(1-\cos
\omega t\right)\coth
\left(\frac{\omega}{2T}\right),
\end{eqnarray}
where $J(\omega)$ is the spectral density, which is defined as $J(\omega)=\sum_{\mathbf{k}}|g_{\mathbf{k}}|^{2}\delta(\omega-\omega_{{\bf
k}})$. In this paper, we consider the Ohmic-family spectral density
\begin{eqnarray}
J(\omega)=\eta\frac{\omega^{s}}{\omega_{c}^{s-1}}e^{-\frac{\omega}{\omega_{c}}},\label{spd}
\end{eqnarray}
where $\omega_{c}$ is the cutoff frequency, $\eta$ is the dimensionless system-bath coupling strength, and $s$ is the Ohmicity
parameter that governs the transition across regimes: sub-Ohmic  ($0<s<1$), Ohmic ($s=1$), and super-Ohmic ($s>1$). Clearly, quantum state ~(\ref{sta1}) is a function of both the above
three descriptive parameters  and time and temperature. In this paper, we will estimate the cutoff frequency $\omega_c$ by measuring quantum state~(\ref{sta1}), and study the relationship between the sensing precision and these variables.

We now employ the quantum parameter estimate theory to quantify the sensing precision for the cutoff frequency $\omega_{c}$.  As established, the sensing precision of $\omega_{c}$ is restricted to the QCR bound
 \begin{equation}
\Delta\omega_{c} \geq \frac{1}{\sqrt{\nu  \mathcal{F}_{q}}}.\label{QCR}
\end{equation}
 Here $\Delta\omega_{c}$ denotes the mean square error, $\nu$ is the number of experimental repetitions and $\mathcal{F}_{q}$ represents the QFI for the $\omega_{c}$. This inequality implies that a larger QFI directly corresponds to tighter bounds and therefore higher potential precision. The QFI only depends on the quantum state of the system under measurement. For state ~(\ref{sta1}), it is given by~\cite{Wei2013,Jing2019}
 \begin{equation}
\mathcal{F}_{q}=\frac{\left(\partial_{\omega_{c}}\gamma\right)^{2}}{\mathrm{e}^{2\gamma}-1},\label{fis}
\end{equation}
where $\partial_{x}Y$ represents the  first-order partial derivative of $Y$ with respect to $x$.

To achieve optimal sensitivity, we identify an observable $\hat{\Lambda}$ that saturates the QCR bound.  In a qubit system, the classical Fisher information for measurement operator
$\hat{X}$ takes the form ~\cite{Mitchison2020}
\begin{equation}
\mathcal{F}_{c}=\frac{\left(\partial_{\omega_{c}}\langle\hat{X}\rangle\right)^{2}}{\langle\Delta \hat{X}^{2}\rangle}\nonumber,
\end{equation}
where $\langle\hat{X}\rangle$ and $\langle\Delta \hat{X}^{2}\rangle$ are its expectation value and variance, respectively.  Crucially, the QFI defines the maximum attainable classical Fisher information across all possible measurements:
\begin{equation}\label{qc}
 \mathcal{F}_{q}=\max_{\hat{X}}\mathcal{F}_{c}=\mathcal{F}_{c}(\hat{\Lambda}).
\end{equation}
As demonstrated in Refs.~\cite{Razavian2019,Yuan2023}, selecting the Pauli-X operator $\hat{\sigma}_{x}$ ($\hat{\Lambda}=\hat{\sigma}_{x}$) establishes equality with~ Eq. (\ref{qc}). Accordingly, we define
$\langle\hat{\sigma}_{x}\rangle$ as our target signal to saturates the QCR bound, which is experimentally accessible through Ramsey interferometry ~\cite{Adam2022,Scelle2013,Cetina2016}.

In our work, we use the dimensionless QSNR
\begin{equation}
\mathcal{Q}=\omega_{c}^{2}\mathcal{F}_{q}\label{Q}
\end{equation}
as a metric for sensing performance. A larger value of $\mathcal{Q}$ indicates higher sensing precision. With respect to the Ohmic spectral density~(\ref{spd}), the QSNR can be further expressed as
\begin{eqnarray}\label{Qs}
\mathcal{Q}(s)=\frac{\left[(1-s)\gamma(s)+\gamma(s+1)\right]^{2}}{e^{2\gamma(s)}-1}.
\end{eqnarray}
Next, based on the Eq.~(\ref{Qs}), we analyze the sensing  performance of $\omega_{c}$ in both zero-temperature and finite-temperature regimes.

\section{\label{Sec:3} Sensing performance at Zero Temperature}
In this section, we investigate the sensing performance of using the qubit as a sensor to estimate Ohmic environment's cutoff frequency at zero temperature. In this case, the decoherence factor $\gamma$ reads~\cite{Breuer2007,Zwerger1987}
\begin{eqnarray}
\label{garm2} \gamma(s) &=\left\{
\begin{array}{c}
\frac{\eta}{2}\ln(1+\omega_{c}^{2}t^{2}), \hspace{0.3 cm} s=1,\vspace{0.3 cm}\\
\eta\left\{1-\frac{\cos\left[(s-1)\arctan(\omega_{c}t)\right]}{(1+\omega_{c}^{2}t^{2})^{\frac{s-1}{2}}}\right\}\Gamma(s-1), \hspace{0.1 cm}
s\neq1,
\end{array}
\right.
\end{eqnarray}
where $\Gamma(s)$ is the Euler gamma function given by the the integral $\Gamma(s)=\int_0^\infty x^{s-1}e^{-x}\mathrm{d}x$. Obviously, substituting the decoherence factor $\gamma$  from Eq.~(\ref{garm2}) into Eq.~(\ref{Qs}) yields an explicit expression for the QSNR. Based on this explicit expression, consider $\omega_{c}t$ as a single variable, we can readily conclude that the QSNR is independent of $\omega_{c}$.

\begin{figure}[tbp]
\includegraphics[width=0.48\textwidth]{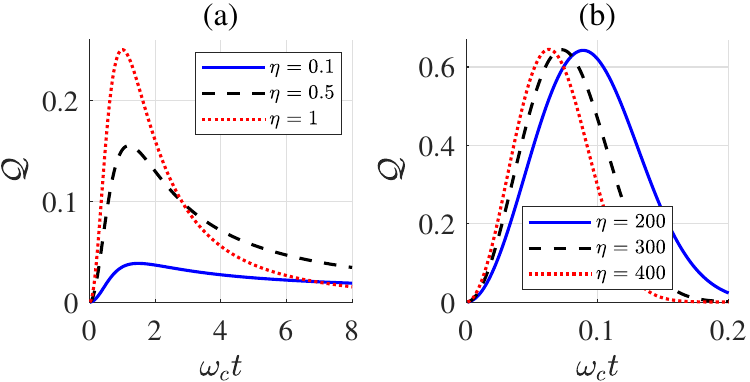}
\caption{(color online) The QSNR versus dimensionless time $\omega_{c}t$ for various $\eta$: (a) small values ($\eta=0.1,0.5,1$); (b) large values ($\eta=200,300,400$), with $s=1$ throughout.} \label{fig1}
\end{figure}
\begin{figure}[tbb]
\includegraphics[clip=true,height=4cm,width=8cm]{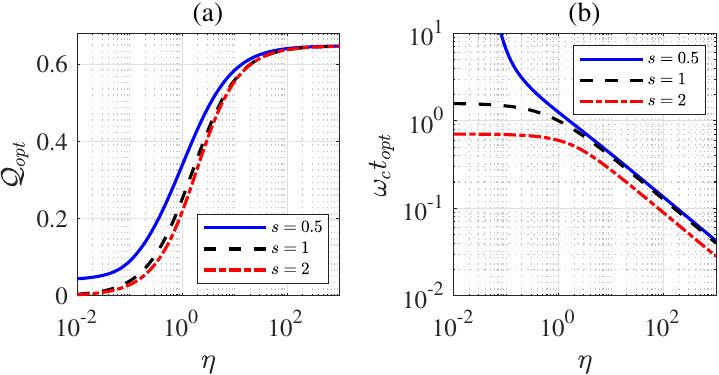}
\caption{(color online)  Dependence of (a) the optimal QSNR and (b) the corresponding optimal measurement time on the coupling strength $\eta$, with the Ohmicity parameter fixed at $s=0.5,1,2$.} \label{fig2}
\end{figure}
\begin{figure}[tbp]
\includegraphics[clip=true,height=4.5cm,width=6cm]{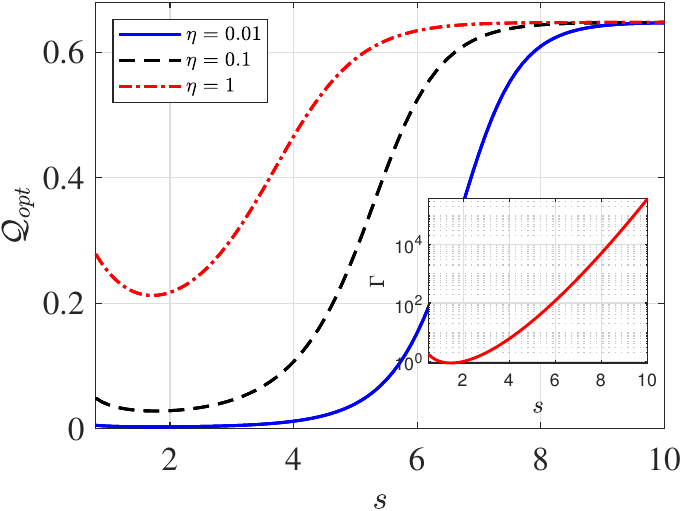}
\caption{(color online) Dependence of the optimal QSNR on the Ohmicity parameter $s$, with the coupling strength fixed at $\eta=0.01,0.1,1$. Inset: Variation of the Euler gamma function with $s$.} \label{fig3}
\end{figure}

As shown in Fig.~\ref{fig1}, the QSNR increases with time as information of the $\omega_{c}$ is gradually incorporated into the decoherence factor. However, due to the reduction in qubit coherence, the QSNR reaches a maximum and subsequently decreases over time. We refer to this peak value as the optimal QSNR $\mathcal{Q}_{\text{opt}}$, and the corresponding time point as the optimal measurement time $t_{\text{opt}}$. For small $\eta$, the optimal QSNR increases with $\eta$, as seen in Fig.~\ref{fig1}(a). But at large $\eta$, it plateaus and becomes almost insensitive to changes in $\eta$, according to Fig.~\ref{fig1}(b).
Since the optimal QSNR depends on both the coupling strength $\eta$ and the Ohmicity parameter $s$, we systematically investigate its behavior across different regimes. As shown in Fig.~\ref{fig2}(a), the optimal QSNR increases monotonically with $\eta$ for fixed $s$, but eventually saturates at approximately 0.65 when $\eta$ becomes sufficiently large regardless of whether the environment is sub-Ohmic ($s=0.5$), Ohmic ($s=1$), or super-Ohmic ($s=2$). Similarly, in Fig.~\ref{fig3}, where the dependence on $s$ is explored for various $\eta$, all curves again converge toward the same limiting value of about $0.65$ as
$s$ increases. Particularly, the red dot-dashed curve for $\eta=1$ in Fig.~\ref{fig3} has already been reported in Ref.~\cite{Benedetti2018}, providing consistency with prior work.


Given that the optimal QSNR keeps leveling off near $0.65$ across different conditions, we are curious about the origin of this limit. In what follows, we aim to reveal the underlying principle behind this phenomenon. As shown in Fig.~\ref{fig3}(b), when the coupling strength $\eta$ becomes sufficiently large, the scaled  optimal measurement time $\omega_{c}t_{opt}$ decreases rapidly with increasing $\eta$. Especially, within the parameter regime where the optimal QSNR reaches its saturation value, the scaled optimal measurement time becomes significantly less than $1$. This distinctive behavior motivates us to investigate how the QSNR evolves over time under the condition $\omega_{c}t\ll1$. Under this condition, we find the decoherence factor in Eq.~(\ref{garm2}) can be simplified to
\begin{eqnarray}
\label{garm3} \gamma=\frac{\eta}{2}\Gamma(s+1)\omega_{c}^{2}t^{2}.
\end{eqnarray}
By applying the identity $\Gamma(s+2)=(s+1)\Gamma(s+1)$, it follows that $\gamma(s+1)=(s+1)\gamma(s)$. As a result, equation~(\ref{Qs}) reduces to
\begin{eqnarray}\label{Qs2}
\mathcal{Q}=\frac{4\gamma^{2}}{\mathrm{e}^{2\gamma}-1}.
\end{eqnarray}
From the extremum condition $\partial_{t}\mathcal{Q}=0$, we obtain a novel transcendental equation
\begin{eqnarray}\label{extr}
\gamma_{\text{opt}}\mathrm{e}^{2\gamma_{\text{opt}}}-\mathrm{e}^{2\gamma_{\text{opt}}}+1=0,
\end{eqnarray}
where $\gamma_{\text{opt}}$ is the optimal decoherence factor at the optimal measurement time $t_{\text{opt}}$. The solution to the above equation yields
\begin{eqnarray}\label{garmopt}
\gamma_{\text{opt}}=1+\frac{1}{2}\mathrm{ProductLog}\left(-\frac{2}{\mathrm{e}^{2}}\right)\approx0.8,
\end{eqnarray}
where the ProductLog function, also known as the Lambert W function or Omega function, is a special function that serves as the inverse of $f(w) = we^{w}$. As a result, the optimal QSNR is found to be
\begin{eqnarray}\label{Qsopt}
\mathcal{Q}_{\text{opt}}=\frac{4\gamma_{\text{opt}}^{2}}{\mathrm{e}^{2\gamma_{\text{opt}}}-1}\approx0.648,
\end{eqnarray}
which is exactly consistent with the numerical results shown in Fig.~\ref{fig2}(a) and Fig.~\ref{fig3}.  We need to point out that, based on both numerical results and analytical analysis, we confirm $0.648$ as the maximum achievable value in our sensing scheme. Therefore,  we refer to the optimal QSNR in Eq.~(\ref{Qsopt}) as the maximum QSNR $$\mathcal{Q}_{\text{max}}=0.648,$$ which is introduced to distinguish it from the optimal QSNR in other parameter regimes. This establishes a criterion for investigating the potential advantages of other schemes in estimating the cutoff frequency of ohmic environments.

Now let us discuss under what conditions the optimal QSNR can reach $\mathcal{Q}_{\text{max}}$. Recall that all results from Eq.~(\ref{garm3}) to Eq.~(\ref{Qsopt}) are all derived under the condition that $\omega_{c}t\ll1$. Therefore, at zero temperature, as long as the scaled optimal measurement time $\omega_{c}t_{\text{opt}}$ satisfies the inequality
\begin{eqnarray}\label{cond1}
\omega_{c}t_{\text{opt}}=\sqrt{\frac{1.6}{\eta\Gamma(s+1)}}<<1,
\end{eqnarray}
regardless of the type of Ohmic spectral density, the optimal QSNR can be saturated to $\mathcal{Q}_{\text{max}}$. To illustrate this point, we consider the behavior of the function $\Gamma(s)$. As shown in the inset of Fig.~\ref{fig3}, $\Gamma$ grows rapidly for sufficiently large $s$. By combining this observation with our derived criterion in Eq.~(\ref{cond1}), we naturally understand why, even for a very small coupling strength such as $\eta = 0.01$, the optimal QSNR can still saturate to $\mathcal{Q}_{\text{max}}$ once $s$ reaches a certain threshold, as illustrated in Fig.~\ref{fig3}. This is because when $\Gamma(s+1)$ is sufficiently large, inequality in Eq.~(\ref{cond1}) can be satisfied even for very small values of $\eta$.

\section{\label{Sec:4} High temperature-enhanced quantum sensing}
In this section, we study how temperature affects the performance of our quantum sensing setup. To provide context, we first highlight two key observations from numerical simulations: (i) even when temperature is present, there's still an optimal measurement time at which the QSNR reaches its peak-meaning, we can still achieve an optimal QSNR; (ii) if we treat the scaled temperature $T/\omega_c$ as a control parameter, then the dependence of the QSNR on the scaled time $\omega_{c}t$ remains entirely independent of the specific value of the cutoff frequency $\omega_{c}$. As seen in Fig.~\ref{fig2}(a) and Fig.~\ref{fig3} from the last section, when both the $\eta$ and $s$ are small in an Ohmic environment, the estimation accuracy is relatively poor. Why? Because these small values prevent the system from reaching the condition where $\omega_{c}t_{\text{opt}}\ll1$, which is necessary for achieving the maximum QSNR $\mathcal{Q}_{\text{max}}$. It's well known that temperature introduces thermal noise, which speeds up decoherence-essentially causing the fragile quantum state to lose information faster. This naturally leads us to ask: what if we turn up the thermal noise enough? Could it actually help push the qubit probe into a regime where the optimal measurement time becomes very short-short enough to satisfy $\omega_{c}t_{\text{opt}}\ll1$? And more importantly, once that happens, what happens to the optimal QSNR? Does it improve, or could  thermal noise ruin everything?
\begin{figure}[tbp]
\includegraphics[width=0.48\textwidth]{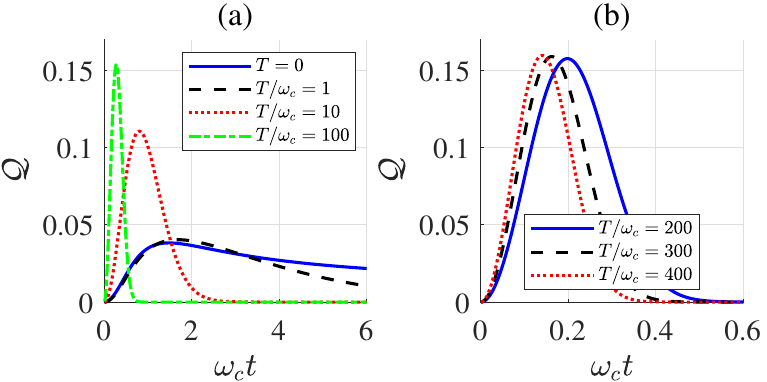}
\caption{(color online) Dependence of the QSNR on the scaled time $\omega_{c}t$ at various  scaled temperatures: (a) $T/\omega_{c}=0,1,10,100$; (b) $T/\omega_{c}=200,300,400$,
 with fixed parameters $s=1$ and $\eta=0.1$.} \label{fig4}
\end{figure}

To address the above questions, we first consider a representative case. As illustrated in Fig.~\ref{fig4}, the dependence of the QSNR on the scaled time $\omega_{c}t$ is plotted for various scaled temperatures $T/\omega_c$, under fixed parameters $s=1$ and $\eta=0.1$. As shown in Fig.~\ref{fig4}(a), increasing the temperature indeed enhances the optimal QSNR, with this enhancement becoming particularly pronounced when $T/\omega_c\gg1$. In this paper, we refer to the temperature regime where $T/\omega_c\gg1$ as "high temperature".  From Fig.~\ref{fig4}(b), we observe that in a high temperature environment, once $T/\omega_c$ exceeds a certain threshold, further increases do not lead to additional growth in the optimal QSNR; instead, it saturates at a constant value. The primary effect of raising $T/\omega_c$ is thus a reduction in the scaled optimal measurement time $\omega_{c}t_{\text{opt}}$. Another notable feature revealed in Fig.~\ref{fig4}(b) is that the scaled optimal measurement time remains significantly less than unity.
\begin{figure}[tbp]
\includegraphics[width=0.48\textwidth]{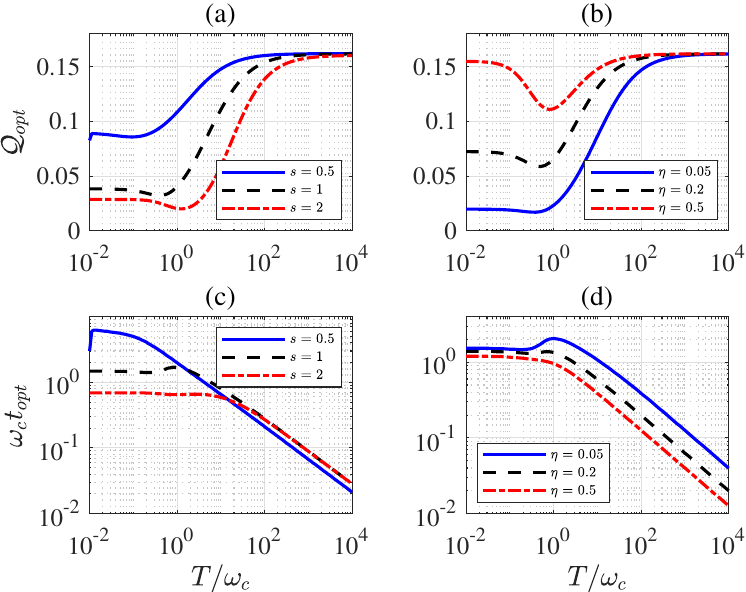}
\caption{(color online) The optimal QSNR (top) and its corresponding optimal measurement time (bottom) as functions of the scaled temperature $T/\omega_{c}$: (left) for $\eta=0.1$ , varying $s=0.5,1,2$; (right) for $s=1$, varying $\eta=0.05,0.2,0.5$.} \label{fig5}
\end{figure}

Having analyzed the representative case, we proceed to examine the general scenario. As shown in Fig.~\ref{fig5}, the optimal QSNR and its corresponding optimal measurement time are plotted as functions of the scaled temperature $T/\omega_c$. In the left panels, we fix $\eta = 0.1$ and vary the Ohmicity parameter $s$ over values 0.5,1, and 2. In the right panels, with $s=1$ held constant, we vary the coupling strength$\eta$ across 0.05, 0.2, and 0.5. Figure~\ref{fig5}(a) demonstrates that when $T/\omega_c<1$, temperature has virtually no effect  on the optimal QSNR. However, for $T/\omega_c>1$, the optimal QSNR increases  monotonically with rising $T/\omega_c$ across all three environments considered. Interestingly, in the high temperature regime, as $T/\omega_c$ continues to increase, all curves converge toward a common saturation value of approximately 0.16. Figure~\ref{fig5}(b) shows that when $\eta=0.05$ and $0.2$ are relatively small, temperature has little effect on the optimal QSNR in the regime where $T/\omega_c<1$. When $\eta=0.5$ is relatively large, increasing temperature slightly reduces the optimal QSNR in the regime around $T/\omega_c\sim1$. Nevertheless,  once $T/\omega_c>1$, all three curves begin to rise with increasing $T/\omega_c$  and ultimately also converge to the same saturation value of about $0.16$ in the high temperature regime. This behavior is similar to the general trend observed in Fig.~\ref{fig5}(a). Such convergence, independent of the spectral density parameters, suggests the possible existence of underlying physical mechanisms driving this phenomenon.

Next, we aim to explain why, when $T/\omega_c\gg1$, the optimal QSNR approaches a constant value of approximately $0.16$ that is independent of the spectral density parameters as $T/\omega_c$ increases. From Figs.~\ref{fig5} (c) and (d), it can be seen that in the high temperature regime, increasing $T/\omega_c$ leads to a rapid decrease in the scaled optimal measurement time. Similar to the zero-temperature case, in the regime where the optimal QSNR saturates, the scaled optimal measurement time also remains much less than 1 ($\omega_{c}t_{\text{opt}}\ll1$). These numerical results motivate us to analyze the behavior of the QSNR under the conditions $T/\omega_c\gg1$ and $\omega_{c}t\ll1$.  When $T/\omega_c\gg1$, it is reasonable to substitute $\coth\left(\omega/2T\right)\approx2T/\omega$ into Eq.~(\ref{garm1}), which then yields
$$\gamma(s,T)=\frac{2T}{\omega_{c}}\gamma(s-1,0).$$ Under the condition $\omega_{c}t\ll1$, based on the result of Eq.~(\ref{garm3}), we directly obtain
\begin{eqnarray}
\label{garm4} \gamma=\eta\Gamma(s)T\omega_{c}t^{2}.
\end{eqnarray}
Based on Eq.~(\ref{garm4}), it is clear that $\gamma(s+1)=s\gamma(s)$. Substituting this result into Eq.~(\ref{Qs}) yields the QSNR as
\begin{eqnarray}\label{Qs3}
\mathcal{Q}=\frac{\gamma^{2}}{\mathrm{e}^{2\gamma}-1}.
\end{eqnarray}
By comparing Eqs.~(\ref{Qs2}) and~(\ref{Qs3}), we find that, for the same value of the decoherence factor, the QSNR under the short-time approximation in the high temperature environment is one-fourth of that in the zero-temperature environment. Clearly, the extremum condition $\partial_{t}\mathcal{Q}=0$ also yields Eq.~(\ref{extr}), and thus leads to the same optimal decoherence factor
$$\gamma_{\text{opt}}=0.8.$$ Then the optimal QSNR under the short-time approximation in the high temperature environment is given by
\begin{eqnarray}\label{Qopt}
\mathcal{Q}_{sat}=\frac{\mathcal{Q}_{\text{max}}}{4}=0.162.
\end{eqnarray}
This value exactly matches the saturation level observed in Figs.~\ref{fig5}(a) and~\ref{fig5}(b). Therefore, to distinguish it from the optimal QSNR in other parameter regimes, we refer to this optimal QSNR as the saturated QSNR, denoted by $\mathcal{Q}_{\text{sat}}$.

We now discuss the conditions under which the optimal QSNR saturates to $\mathcal{Q}_{\text{sat}}$. Through the above analysis, we find that for the optimal QSNR to saturate at $\mathcal{Q}_{\text{sat}}$, two conditions must be satisfied: (i) The Ohmic environment is in a high temperature regime ($T/\omega_c\gg1$). (ii) The optimal measurement time $t_{\text{opt}}$ corresponding to reaching
$\mathcal{Q}_{\text{sat}}$ is sufficiently short ($\omega_{c}t_{\text{opt}}\ll1$). By  substituting
$\gamma_{\text{opt}}=0.8$ for $\gamma$ in Eq.~(\ref{garm4}), the scaled optimal measurement time $\omega_{c}t_{\text{opt}}$ is given by
\begin{eqnarray}
\label{optt3} \omega_{c}t_{\text{opt}}=\sqrt{\frac{0.8\omega_{c}}{\eta\Gamma(s)T}}.
\end{eqnarray}
Equation~(\ref{optt3}) shows that for any Ohmic environment, as long as $T/\omega_c$ is sufficiently large, one can achieve $\omega_{c}t_{\text{opt}}\ll1$, thereby enabling the optimal QSNR to reach
$\mathcal{Q}_{\text{sat}}$. Ultimately, the condition for the optimal QSNR to saturate at
$\mathcal{Q}_{\text{sat}}$ can be reduced to: the scaled temperature $T/\omega_c$ is much greater than $1$, such that $\omega_{c}t_{\text{opt}}\ll1$.

Figures~\ref{fig2}(a) and~\ref{fig3} indicate that, at zero temperature, the optimal QSNR is significantly smaller than
$\mathcal{Q}_{\text{max}}$ in Ohmic environments with relatively small $\eta$ and $s$. We also conclude that sufficiently high temperature can always drive the optimal QSNR to reach
$\mathcal{Q}_{\text{max}}/4$, irrespective of the specific Ohmic environment. This implies that for Ohmic environments characterized by small $\eta$ and $s$, we can improve the precision of the sensing cutoff frequency by introducing sufficiently high temperature. To quantify this enhancement effect, we introduce the high temperature enhancement factor $R$, defined as
\begin{eqnarray}
R=\frac{\mathcal{Q}_{\text{sat}}}{\mathcal{Q}_{\text{opt}}(T=0)}.
\end{eqnarray}
This ratio quantifies how much the sensing precision is enhanced due to high temperature. A value of
$R>1$ indicates that high temperature improves the sensor's performance compared to the zero-temperature case. The larger the value of
$R$, the more significant the enhancement effect.
\begin{figure}[tbp]
\includegraphics[width=0.48\textwidth]{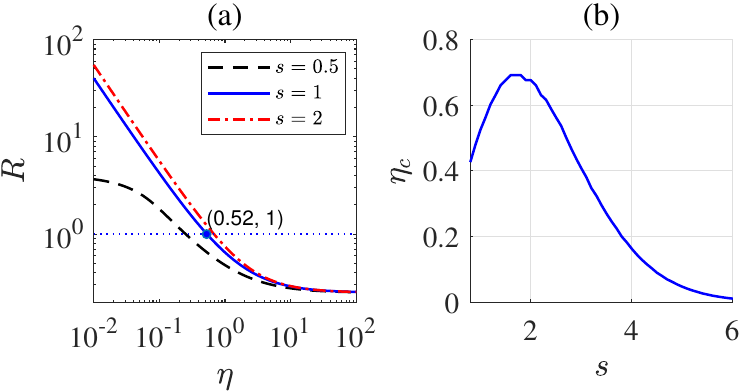}
\caption{(color online) (a) Dependence of the high temperature enhancement factor $R$ on the coupling strength $\eta$, with the Ohmicity parameter fixed at $s=0.5,1,2$. (b) Variation of the critical coupling strength $\eta_{c}$ as a function of the Ohmicity parameter $s$.  } \label{fig6}
\end{figure}
Figure~\ref{fig6}(a) shows the dependence of the high temperature enhancement factor $R$ on the coupling strength $\eta$, with the Ohmicity parameter fixed at $s=0.5, 1,$ and $2$. The plot reveals that in all three environments, the enhancement effect of high temperature on sensing precision becomes increasingly pronounced as $\eta$ decreases. When $\eta$ is on the order of $10^{-2}$, the optimal QSNR increases by nearly two orders of magnitude compared to its zero-temperature value in both Ohmic ($s=1$) and super-Ohmic ($s=2$) cases; even for the sub-Ohmic environment ($s=0.5$), a several-fold improvement is observed. Furthermore, we find that for each fixed value of $s$, there exists a critical coupling strength $\eta_{c}$-for instance, $\eta_{c} = 0.52$ when $s=1$-such that high temperature enhances the sensing precision only when $\eta < \eta_{c}$. The variation of the critical coupling strength $\eta_{c}$ with the Ohmicity parameter $s$ is shown in Fig.~\ref{fig6}(b). The plot reveals that for all values of $s$, $\eta_{c}<1$. This explains why prior studies on the cutoff frequency sensing in Ohmic environments with $\eta=1$-which lies above the critical threshold-did not observe enhancement of sensing precision through high temperature\cite{Sehdaran2019}.

\section{\label{Sec:5} conclusion}
 We investigated the quantum sensing performance of using a single dephasing qubit as a probe to estimate the cutoff frequency in Ohmic environments. The performance is quantified by the QSNR $\mathcal{Q}$. We show that the evolution of the QSNR with scaled time $\omega_c t$ is independent of $\omega_c$ itself, and for fixed $\eta$ and $s$, it always reaches a maximum during the dynamics. Under zero-temperature conditions, we examined how the optimal QSNR depends on $\eta$ and $s$. It was found that as $\eta$ and $s$ increase, the optimal QSNR increases monotonically until saturating at $\mathcal{Q}_{\text{max}} = 0.648$, which represents the maximum achievable value within our sensing scheme. We analytically derived this saturation value and identified the condition for its attainment: $\omega_c t_{\text{opt}} = \sqrt{1.6/(\eta\Gamma(s+1))} \ll 1$. For Ohmic environments with relatively small $\eta$ and $s$, this condition is not satisfied. Numerical results confirmed that, under such circumstances, the optimal QSNR falls significantly below $\mathcal{Q}_{\text{max}}$.

We further studied the influence of temperature on sensing performance. In the high-temperature regime, provided that $\omega_c t_{\text{opt}} = \sqrt{0.8\omega_c/(\eta\Gamma(s)T)} \ll 1$, the optimal QSNR consistently reaches $\mathcal{Q}_{\text{sat}} = \mathcal{Q}_{\text{max}}/4 = 0.162$, regardless of the specific type of Ohmic environment. This indicates that for environments with small $\eta$ and $s$, where zero-temperature precision is limited, introducing sufficiently high temperature can enhance the estimation accuracy of the cutoff frequency. To quantify this enhancement, we introduced the high-temperature enhancement factor $R$ and analyzed its dependence on $\eta$ for fixed values of $s = 0.5, 1,$ and $2$. The results show that in all three cases, the beneficial effect of high temperature becomes more pronounced as $\eta$ decreases. At $\eta \sim 10^{-2}$, the optimal QSNR improves by nearly two orders of magnitude compared to the zero-temperature case in both Ohmic ($s=1$) and super-Ohmic ($s=2$) environments, and still increases several-fold even in the sub-Ohmic case ($s=0.5$). Finally, we identified that for each fixed $s$, there exists a critical coupling strength $\eta_c$: only when $\eta < \eta_c$ does high temperature exert a positive impact on sensing precision. The relationship between $\eta_c$ and $s$ is also presented.

Traditionally, it is believed that temperature makes the coherence of qubits harder to maintain, thereby degrading quantum sensing performance. This conclusion is clearly unquestionable within sensing frameworks based on stationary states of qubits. However, our study focuses on sensing performance utilizing dynamical states of a qubit probe, revealing a counterintuitive phenomenon: under specific conditions, an increase in temperature not only fails to degrade but can actually enhance the precision of parameter estimation. The origin of this seemingly paradoxical beneficial effect lies in the fact that elevated temperatures facilitate a faster encoding rate of environmental information into the dynamical state of the qubit. This effectively compresses the optimal measurement time into a short-time regime, significantly boosting sensing precision. Consequently, our findings suggest that, in certain scenarios, temperature should not be viewed merely as a source of disturbance, but rather as a valuable resource. Indeed, temperature has  recently demonstrated positive effects in other quantum sensing processes~\cite{Ostermann2024,Yuan2025}. By accelerating information acquisition to achieve higher precision, our work provides a novel perspective for realizing enhanced quantum sensing performance.

\acknowledgments
J. B.  Yuan was supported by  NSFC (No. 11905053), the Natural Science Foundation of Hunan Province (Grant
No. 2025JJ50005), and the Open Fund of the Innovation Platform for Micro-Nano Energy Materials and Application Technology, Hunan Provincial Key Laboratory of Higher Education Institutions (Grant No. 2025HSKFJJ016). L. M. Kuang  was supported by NSFC (Grant Nos. 12175060, 12247105, 12421005), Hunan Provincial Major Sci-Tech Program (Grant No. 2023ZJ1010), Henan Science and Technology Major Project (No. 241100210400), and  XJ-Lab key project (Grant No. 23XJ02001). Y. J. Song was supported by  NSFC (No. 12205088). S. Q. Tang was supported by Scientific Research Fund of Hunan Provincial Education Department of China under Grant (No. 22A0507).

\end{document}